\pdfoutput=1
\documentclass[12 pt]{article}
\usepackage{graphicx}
\usepackage{amssymb}
\usepackage{amsmath}
\usepackage{amsfonts}
\usepackage{amsthm}
\usepackage[english]{babel}
\usepackage[latin1,ansinew]{inputenc}
\usepackage{a4wide}
\usepackage{color}
\newcommand{\be}{\begin{eqnarray}}
\newcommand{\ee}{\end{eqnarray}}

\begin{document}
\date{}
\title{\textbf{Jump relations for magnetrohydrodynamic shock waves in a dusty gas atmosphere}}
\author{\textbf{Raj Kumar Anand} \\Department of Physics, UGC Centre of Advanced Studies
\\University of Allahabad, Prayagraj 211002, India
\\ \textit{email: rkanand@allduniv.ac.in, anand.rajkumar@rediffmail.com}}




\maketitle

%

\begin{abstract}
In this article, we have derived Rankine-Hugoniot (RH) jump conditions across a magnetohydrodynamic (MHD) shock front propagating in a dusty gas atmosphere. The dusty gas atmosphere is assumed to be a mixture of a perfect gas and small solid particles, in which small spherical solid particles are continuously distributed. The non-relativistic RH conditions for the pressure, the density, and the fluid velocity across an MHD shock front have been derived in terms of a compression ratio. The simplified forms of RH conditions have been written simultaneously for the weak and strong MHD shock waves in terms of the initial volume fraction of solid particles, the ratio of specific heats of the mixture, and the strength of the magnetic field. Further, the weak and strong shocks have been explored under two distinct conditions, viz., (i) when the applied magnetic field is weak and (ii) when the field is strong. Finally, the effects on the shock velocity and the pressure across the MHD shock front are studied due to the strength of the magnetic field, the concentration of dust particles in the mixture, and the volumetric parameter. This study presents an overview of the influence of the strength of the magnetic field and the dust loading parameters on the shock velocity, the pressure, the density, and the fluid velocity across the MHD shock front. 

\end{abstract}

\section{Introduction}
\label{intro}
The high-speed flows occur in natural phenomena and contain shock waves or blast waves. The supersonic motion has a strong tendency to cause shock waves. For example, hot stars produce winds that move highly supersonically into the interstellar medium (ISM) and produce shock waves \cite{Miles2009}. Similarly, the galaxies in a cluster move supersonically through the cluster gas, heating it via shocks. The gravitational or rotational forces can accelerate gas to supersonic speeds or accelerate objects to move through gas supersonically. Any deceleration or deflection of supersonic gas inevitably leads to shock waves, as when gas falls near radially down magnetic field lines onto a neutron star. The magnetic field strengths encountered in astrophysics range from $10^{-6}$ G in the hot ISM to $10^{12}$ G on the surface of a neutron star. The magnetic fields have a significant effect on the dynamics of astrophysical fluids \cite{Inoue2009}. 

The Rankine-Hugoniot (RH) conditions are one of the foundations of modern shock physics and quickly provide data of shock parameter values for many practical applications rather than obtaining more precise data by elaborately solving hyperbolic differential equations. These jump conditions have contributed greatly to the application and analysis of shock waves in several branches of science and engineering and have turned shock wave physics into an interdisciplinary field. The RH conditions are widely applied in scientific research, such as in sonic booms, supersonic aerodynamics, detonation physics, high-speed combustion, geophysics, astrophysics, fluid and gas dynamics, materials science, relativity theory, quantum mechanics, traffic flow analysis, ocean dynamics, meteorology, magnetohydrodynamics, cosmical gas dynamics, and computational fluid dynamics. The shock parameters of all aggregation states and dimensions, from laboratory to cosmic dimensions, have been determined by experimentalists using these RH conditions.

The magnetohydrodynamic (MHD) shock waves are created due to reconnection processes \cite{Priest2008} as occur in many astrophysical outflows such as extragalactic relativistic jets \cite{Appl1988,Keppens2008}. These MHD shock waves accelerate particles to relativistic speeds \cite{Kirk2004}.  The study of effects on the shock waves due to the presence of small solid particles and magnetic fields has been a motivation, as it is academically an important part of magnetohydrodynamics. The jump conditions across an MHD shock have great scientific importance in many problems that belong to astrophysics, space, and earth sciences.  Hoffmann and Teller \cite{Hoffmann1950} provided a mathematical treatment for the MHD shock waves in the extreme cases of very weak and very strong magnetic fields. Whitham \cite{Whitham1958} investigated the motion of MHD shocks in an ideal gas. Bazer and Ericson \cite{Bazer1959} were among the first to explore the astrophysical applications of the hydromagnetic shock waves. Feng-si \cite{Feng1984} obtained the dimensionless jump conditions for MHD shock waves in a perfect gas, and Anand \cite{Anand2013} derived the jump conditions for MHD shocks propagating in a non-ideal gas. The shock or blast waves in dusty gas have drawn the attention of   Carrier \cite{Carrier1958}, Kribel \cite{Kribel1964}, Marble \cite{Marble1970}, Outa et al. \cite{Outa1976}, Higashino and Suzuki \cite{Higashino1980}, Rudinger \cite{Rudinger1980}, Pai et al. \cite{Pai1980}, Miura and Glass \cite{Miura1985}, Igra and Ben-Dor \cite{Igra1988}, Ben-Dor \cite{Ben-Dor1996}, Conforto \cite{Conforto2001}, Steiner and Hirschler \cite{Steiner2002}, Saito et al. \cite{Saito2003}, Gretler and Regenfelder  \cite{Gretler2008}, Laibe and Price \cite{Laibe2014}, Anand \cite{Anand2014}, Gupta et al. \cite{Gupta2016}, Frost \cite{Frost2018}, and others. The shock or blast waves through dusty gas in the presence of magnetic fields have been studied by Nath \cite{Nath2015,Nath2017}, Vishwakarma et al. \cite{Vishwakarma2018}, Sharma et al. \cite{Sharma2019}, Chaudhary and Singh \cite{Chaudhary2019}, Sahu \cite{Sahu2020}, Pang et al. \cite{Pang2020}, and others. The aim of writing this article is to present the simplified forms of RH jump conditions across an MHD shock front in a dusty gas atmosphere, required for pursuing research of shock or blast waves in dusty magnetohydrodynamics. 

In this article, we have derived, for the first time, RH jump conditions (non-relativistic) for MHD shock waves propagating in a two-phase gas-particle atmosphere. In view of the Pai model  \cite{Pai1980,Pai1977,Pai1991}, the dusty gas is a two-phase mixture of a perfect gas and small solid particles, in which solid particles are spherical and continuously distributed. The diameter of the particle is much smaller than the characteristic length of flow field. The RH jump conditions for the pressure, the density, and the fluid velocity across an MHD shock front have been derived in terms of a single parameter $\xi$ (compression ratio), which characterizes the strength of the shock wave. The handy forms of these RH conditions have been simplified and written in terms of the initial volume fraction of solid particles, the ratio of specific heats of the mixture, and the strength of the magnetic field, simultaneously for the weak and strong MHD shock waves. Further, the weak and strong MHD shocks are explored under two distinct conditions, viz., (i) when the applied magnetic field is weak and (ii) when the field is strong. Finally, the effects due to (i) the concentration of solid particles in the mixture, (ii) the volumetric parameter, i.e., the ratio of the density of solid particles to the initial density of gas, and (iii) the strength of the magnetic field have been investigated on the shock velocity and the pressure across the MHD shock front. It is worth mentioning that RH jump conditions across an MHD shock in a two-phase gas-particle medium reduce to the well-known RH conditions for MHD shocks in ideal gas \cite{Whitham1958} when the mass fraction (concentration) of solid particles in the mixture becomes zero. Thus, the results provided a clear picture of whether and how the presence of small solid particles and magnetic fields affects the state variables behind the MHD shock front. 

The structure of this paper is organized as follows: In the next Sect. \ref{conditions} we present the construction of RH conditions for MHD shock waves. Section \ref{results} contains the analysis with discussion on important components. Concluding remarks are given in Sect. \ref{concl}. 

\section{RH jump conditions across an MHD shock front}
\label{conditions}
In this section, RH jump conditions are derived for weak and strong MHD shock waves propagating in a dusty gas atmosphere. The dusty gas is assumed to be a homogeneous mixture of a perfect gas and small solid particles having initially uniform distribution of density, and it is flowing in the x-direction only (for detail see \cite{Conforto2001,Gretler2008,Anand2014} and the references therein). The equation of state of the mixture of ideal gas and small solid particles under equilibrium conditions is $p=\dfrac{1-k_p}{1-z}\rho R_i T$, where $k_p$ is the mass fraction (concentration) of solid particles in the mixture, $z$ is the volumetric fraction of solid particles in the mixture, $R_i$ is the gas constant, and $T$ is the temperature of the mixture. The relation between $k_p$ and $z$ is given by Pai et al. \cite{Pai1980} as follows: $k_p=z\rho_{sp}/\rho$, where $z=z_o\rho/\rho_o$, while $\rho_{sp}$ is the species density of the solid particles and a subscript $o$ refers to the initial values of $z$ and $\rho$. The concentration of solid particles $k_p$ is defined as $k_p=m_{sp}/m_g$, where $m_{sp}$ is the total mass of solid particles, and $m_g$ is the total mass of the mixture. It is notable that in equilibrium flow, the mass concentration of solid particles $k_p$ remains uniform in the whole flow field.  
 
The internal energy $e$ per unit mass of the mixture is defined as $e=(1-z)p/(\Gamma-1)\rho$, where $z$, the volumetric fraction, is defined as $z=V_{sp}/V_g$, and $\Gamma$ is the ratio of the specific heats of the mixture given by $\Gamma=\left[\gamma(1-k_p)+k_p \beta_{sp}\right]/ \left[1-k_p+k_p \beta_{sp}\right]$. Here, $V_{sp}$ is the volumetric extension, $V_g$ is the total volume of the mixture, $\gamma = c_p/c_v$ is the ratio of specific heats of the gas, and $\beta_{sp}$ is the ratio of specific heats of the solid particles. The initial volume fraction of solid particles is given by $z_o=k_p/\left[G(1-k_p)+k_p\right]$, where $G$ is the volumetric parameter, i.e., the ratio of the density of solid particles $\rho_{sp}$ to the initial density of the gas $\rho_g$. Thus, the fundamental parameters of the Pai model \cite{Pai1980,Pai1977,Pai1991} are $k_p$ and $G$, which describe the effects of the dust loading. For the dust-loading parameter $G$, we have a range of $G=1$ to $G\longrightarrow\infty$, i.e., $V_{sp}\longrightarrow 0$.
  
The shock wave is a single unsteady wavefront with no thickness or a single steady wave of finite thickness. The thickness of shock waves is about $10^{-7}$ \textit{m} in air at ambient conditions, and they arise due to the deposition of large amounts of energy in a very small region over short intervals. The RH conditions are a set of equations relating the state variables of the shocked medium to the ones of the undisturbed medium. If the magnetic field is perpendicular to the shock front, then the flow is entirely along the magnetic field lines and remains unaffected by the magnetic field. Therefore, in such a case, the RH conditions are the same as in the non-magnetic case \cite{Anand2014}. If the magnetic field is parallel to the shock front, then we need to include the magnetic terms in the momentum equation and the energy equation. Here it is assumed that the magnetic field is uniform in the upstream region. 

If $r=R(t)$ is the position of the shock front, then the velocity of the shock front is given by $U= \frac{dR}{dt}$. The RH conditions are obtained from the following principles of conservation of mass, magnetic flux, momentum, and energy: 
\begin{eqnarray}\label{eq-1}
\rho(U-u)=\rho_o U, H(U-u)=H_o U,\nonumber \\
p+\frac{\mu H^{2}}{2}+\rho(U-u)^2=p_o+\frac{\mu H_{o}^{2}}{2}+\rho_o U^2,\\
e+\frac{p}{\rho}+\frac{(U-u)^2}{2}+\frac{\mu H^2}{\rho}=e_o+\frac{p_o}{\rho_o}+\frac{U^2}{2}+\frac{\mu H_o^2}{\rho_o}.\nonumber
\end{eqnarray}
where $u$ is the velocity of the mixture, $\rho$ is the density of the mixture, $p$ is the pressure of the mixture, $\mu$ is the magnetic permeability of the mixture, $H$ is the magnetic field, and $e$ is the internal energy per unit mass of the mixture. In the equilibrium state, the quantities without a suffix refer to the quantities behind the shock front, whereas the quantities with a suffix $o$ refer to the quantities ahead of the shock front. The shock jump conditions (\ref{eq-1}) are also valid for a curved shock front, e.g., in a spherical medium, because the thickness of the shock front is almost always negligible compared to its radius of curvature.

If we take $\xi=\rho/\rho_o$ as a parameter characterizing the shock strength, then equation (\ref{eq-1}) representing MHD shock conditions may be written as: 
\begin{eqnarray}\label{eq-2}
\rho&=&\rho_o \xi, H=H_o \xi, u= U(\xi-1)/\xi, \nonumber\\
U^2&=&\frac{2\xi}{(\Gamma+1)-(\Gamma-1+2z_o)\xi}\left[a_o^2(1-z_o)+b_o^2\left\lbrace\left(1-\frac{\Gamma+z_o(\xi+1)}{2}\right)\xi+\frac{\Gamma}{2}\right\rbrace\right],\\
p&=&p_o+\frac{2\rho_o(\xi-1)}{(\Gamma+1)-(\Gamma-1+2z_o)\xi}\left[a_o^2(1-z_o)+\frac{b_o^2(\Gamma-1)(\xi-1)^2}{4} \right].\nonumber
\end{eqnarray}
where $a_o=\left[\Gamma p_o/(1-z_o)\rho_o\right]^{1/2}$ is the speed of sound in an unperturbed medium, and $b_o=\left(\mu H_o^2/\rho_o\right)^{1/2}$ is the Alfven speed.  The above shock jump conditions (\ref{eq-2}) reduce to the well-known RH shock conditions \cite{Whitham1958} for shock waves in an ideal gas when $z_o$, the initial volume fraction of small solid particles, becomes zero. The strength of the magnetic field \cite{Anand2013} is measured by the ratio of the Alfven speed to the speed of sound in an unperturbed medium,  i.e., $b_o^2/a_o^2=\beta^2$ (say). 

\subsection{Weak MHD shock waves}
In the limiting case of weak shocks, $p/p_o$ is very small. The parameter $\xi$ is slightly greater than unity. Therefore, we may write $\xi=1+\epsilon$, where $\epsilon$ is another parameter that is negligible in comparison with unity, i.e., $\epsilon\ll1$.
\subsubsection{Weak shock with weak magnetic field} 
\label{wswmf} 
When $b_o^2 \ll a_o^2$, or $\mu H_o^2 \ll p_o$, i.e., the magnetic pressure is very much smaller than the pressure of the mixture, then the MHD shock jump conditions (\ref{eq-2}) become 
\begin{eqnarray}\label{eq-3}
\rho&=&\rho_o(1+\epsilon), H=H_o(1+\epsilon), u= U\epsilon, \nonumber\\
\frac{U}{a_o}&=& 1+\frac{(\Gamma+1)+3(1-z_o)\beta^2}{4(1-z_o)}\epsilon,\\
\frac{p}{p_o}&=& 1+\frac{\Gamma}{1-z_o}\epsilon .\nonumber
\end{eqnarray}
Equation (\ref{eq-3}) represents the handy form of RH conditions for weak shocks in the presence of a weak magnetic field. 
\subsubsection{Weak shock with strong magnetic field}
\label{wssmf} 
When $b_o^2 \gg a_o^2$, or $\mu H_o^2 \gg p_o$, i.e., the magnetic pressure is very much larger than the pressure of the mixture, then the MHD shock jump conditions (\ref{eq-2}) become 
\begin{eqnarray}\label{eq-4}
\rho&=&\rho_o(1+\epsilon), H=H_o(1+\epsilon), u= U\epsilon, \nonumber\\
\frac{U}{b_o}&=& 1+\frac{(\Gamma+1)+3(1-z_o)\beta^2}{4(1-z_o)\beta^2}\epsilon,\\
\frac{p}{p_o}&=& 1+\frac{\Gamma}{1-z_o}\epsilon .\nonumber
\end{eqnarray}
Equation (\ref{eq-4}) represents the handy form of RH conditions for weak shocks in the presence of a strong magnetic field. 
\subsection{Strong MHD shock waves}
In the limiting case of strong shock waves, $p/p_o$ is large. In the magnetic case, this is achieved in two ways:(i) the purely non-magnetic way, when $\xi$ is close to the value of $(\Gamma+1)/(\Gamma-1+2z_o)$, and (ii) when $b_o^2 \gg a_o^2$ or $\mu H_o^2 \gg p_o$, the magnetic pressure is very much greater than the pressure of the mixture in the equilibrium state.
\subsubsection{Strong shock with weak magnetic field} 
\label{sswmf} 
When $b_o^2 \ll a_o^2$, or $\mu H_o^2 \ll p_o$, i.e., the magnetic pressure is very much smaller than the pressure of the mixture, then the MHD shock jump conditions (\ref{eq-2}) become 
\begin{eqnarray}\label{eq-5}
\rho&=&\rho_o\xi, H=H_o\xi, u= U(\xi-1)/\xi, \nonumber\\
p/p_o&=&1+\left(\chi^{'}a_o^2+A^{'}b_o^2\right)U^2/a_o^4,
\end{eqnarray}
where $\chi^{'}=\frac{\Gamma(\xi-1)}{(1-z_o)\xi}$ and $A^{'}=\frac{\Gamma(\xi-1)\left[(\Gamma-1)(\xi-1)^2-2\lbrace(2-z_o(\xi+1)-\Gamma)\xi+\Gamma\rbrace\right]}{4(1-z_o)^2\xi}$.

Equation (\ref{eq-5}) represents a handy form of RH conditions for strong shock waves in the presence of a weak magnetic field.
\subsubsection{Strong shock with strong magnetic field} 
\label{sswmf} 
When $b_o^2 \gg a_o^2$, or $\mu H_o^2 \gg p_o$, i.e., the magnetic pressure is very much larger than the pressure of the mixture, then the MHD shock jump conditions (\ref{eq-2}) become 
\begin{eqnarray}\label{eq-6}
\rho&=&\rho_o\xi, H=H_o\xi, u= U(\xi-1)/\xi, \nonumber\\
p/p_o&=&1+\chi\left(b_o^2+A a_o^2\right)U^2/a_o^2 b_o^2,
\end{eqnarray}
where $\chi=\frac{\Gamma(\xi-1)^3(\Gamma-1)}{2(1-z_o)\xi\lbrace(2-z_o(\xi+1)-\Gamma)\xi+\Gamma\rbrace}$ and $A=\frac{4(1-z_o)}{(\Gamma-1)(\xi-1)^2}-\frac{2(1-z_o)}{(2-z_o(\xi+1)-\Gamma)\xi+\Gamma}$.

Equation (\ref{eq-6}) represents a handy form of RH conditions for strong shock waves in the presence of a strong magnetic field.

\section{Results and Discussion}
\label{results}
This section presents an analysis of RH jump conditions derived for one-dimensional MHD shock waves propagating in a dusty gas atmosphere consisting of an ideal gas and spherically small solid particles. It is worth mentioning that these RH shock conditions are valid for the exploding and imploding MHD shock waves and reduce to the well-known RH conditions \cite{Whitham1958} for the MHD shocks propagating in an ideal gas when $z_o$, the initial volume fraction of small solid particles, becomes zero. The $z_o$ is a function of $k_p$, the mass concentration of the solid particles, and $G$, the ratio of the density of solid particles to the initial density of gas. The strength of the magnetic field is measured by the ratio of Alfven speed to sound speed in the medium ahead of the shock front, i.e., $b_o^2/a_o^2=\beta^2$ (say). The typical values of parameters are taken as $2<\xi<4$, $0<\beta^2<20$, $\gamma =7/5$, $\beta_{sp}=1$, $k_p = 0, 0.01, 0.02, 0.2, 0.4$, and $G = 1, 10, 100, \infty$, for numerical computation of shock velocity and pressure using Mathematica-8. The value of $\beta_{sp}=1$ and $\gamma=7/5$ corresponds to the mixture of air and glass particles \cite{Miura1985}. In present analysis, we have assumed the initial volume fraction of solid particles $z_o$ to be a small constant. Obviously, the value $\beta^2=0$ corresponds to a non-magnetic case \cite{Anand2014}, however, $\beta^2>0$ corresponds to a magnetic case. The value $k_p=0$ corresponds to the case of a dust-free gas. The influence of weak and strong magnetic fields on the shock velocity and the pressure across the shock front has been investigated, respectively, for the weak and the strong MHD shock waves in a two-phase gas-particle atmosphere. 
\subsection{Weak MHD shock waves}
Now we explore the influence of the magnetic field on the weak shock waves propagating in a two-phase gas-particle atmosphere under two conditions viz., (i) when the magnetic field is weak and (ii) when the field is strong, respectively.

\begin{figure}   
   \begin{minipage}{0.4\textwidth}
     \centering
     \includegraphics[width=1.2\linewidth, trim=0 .1cm .1cm 0, clip]{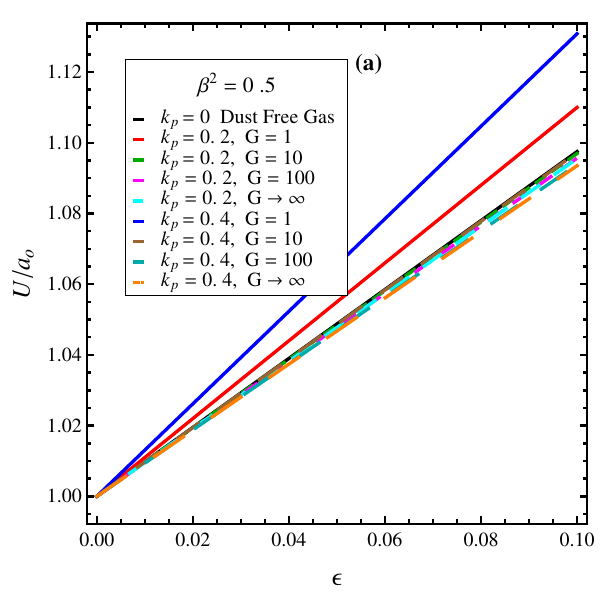}
       \end{minipage}\hfill
   \begin{minipage}{0.4\textwidth}
     \centering
     \includegraphics[width=1.2\linewidth, trim=0 .1cm .1cm 0, clip]{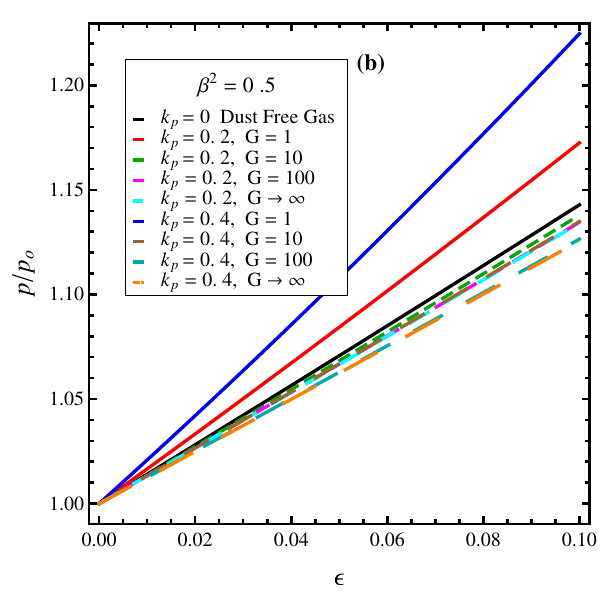}
     \end{minipage}
     \begin{minipage}{0.4\textwidth}
     \centering
     \includegraphics[width=1.2\linewidth, trim=0 .1cm .1cm 0, clip]{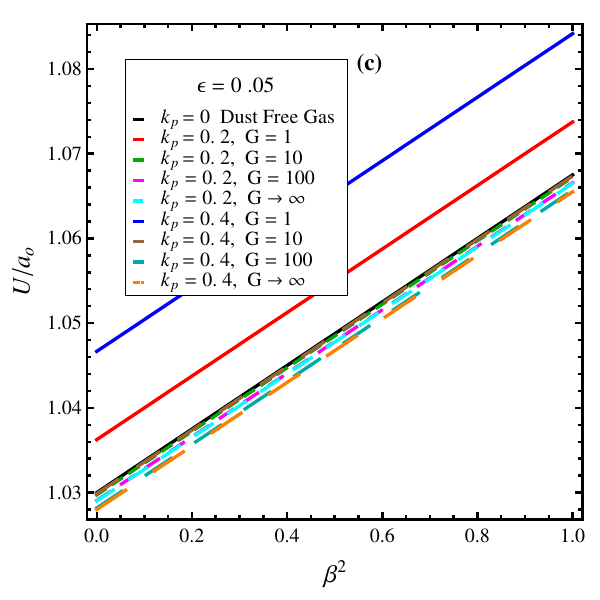}
       \end{minipage}\hfill
   \begin{minipage}{0.4\textwidth}
     \centering
     \includegraphics[width=1.2\linewidth, trim=0 .1cm .1cm 0, clip]{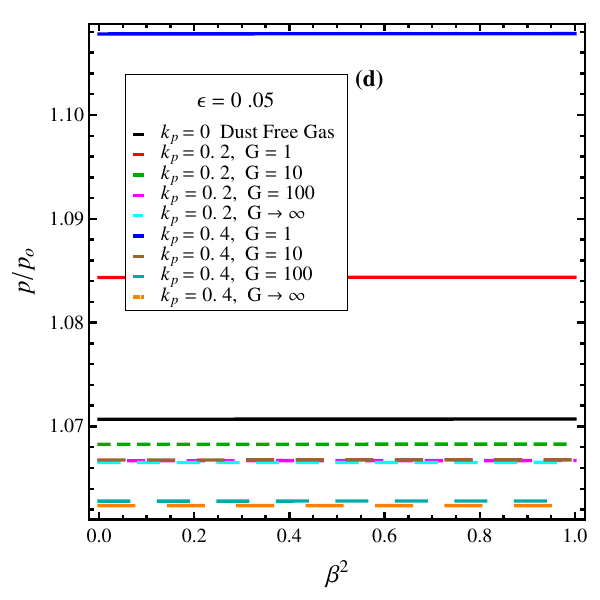}
     \end{minipage}
     \caption{Variations of (a) $U/a_{o}$ vs. $\xi$, (b) $p/p_{o}$ vs. $\xi$, (c) $U/a_{o}$ vs. $\beta^2$,  (d) $p/p_{o}$ vs. $\beta^2$ in case of weak MHD shock waves in the presence of a weak magnetic field}\label{figure1a_1d}
\end{figure}
\begin{figure}   
   \begin{minipage}{0.4\textwidth}
     \centering
     \includegraphics[width=1.2\linewidth, trim=0 .1cm .1cm 0, clip]{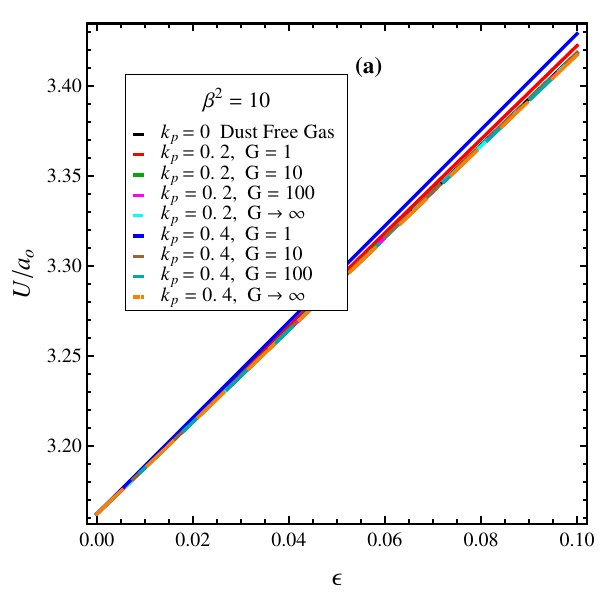}
       \end{minipage}\hfill
   \begin{minipage}{0.4\textwidth}
     \centering
     \includegraphics[width=1.2\linewidth, trim=0 .1cm .1cm 0, clip]{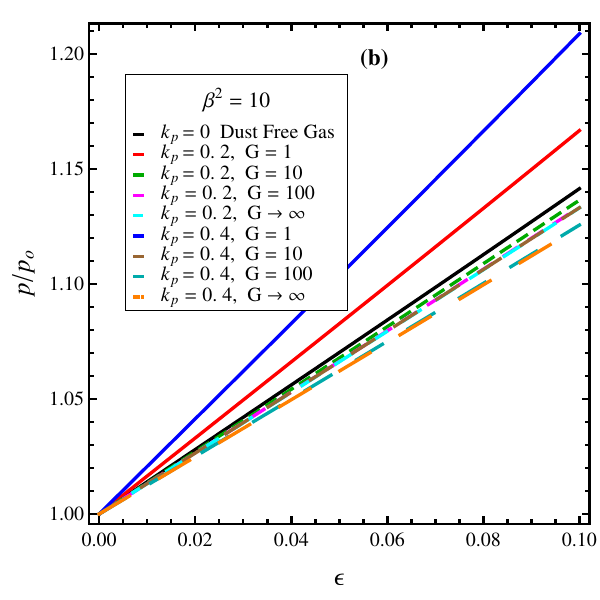}
     \end{minipage}
     \begin{minipage}{0.4\textwidth}
     \centering
     \includegraphics[width=1.2\linewidth, trim=0 .1cm .1cm 0, clip]{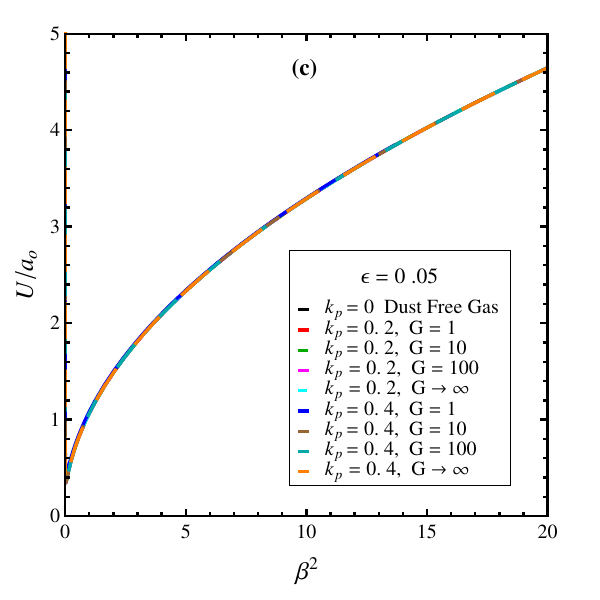}
       \end{minipage}\hfill
   \begin{minipage}{0.4\textwidth}
     \centering
     \includegraphics[width=1.2\linewidth, trim=0 .1cm .1cm 0, clip]{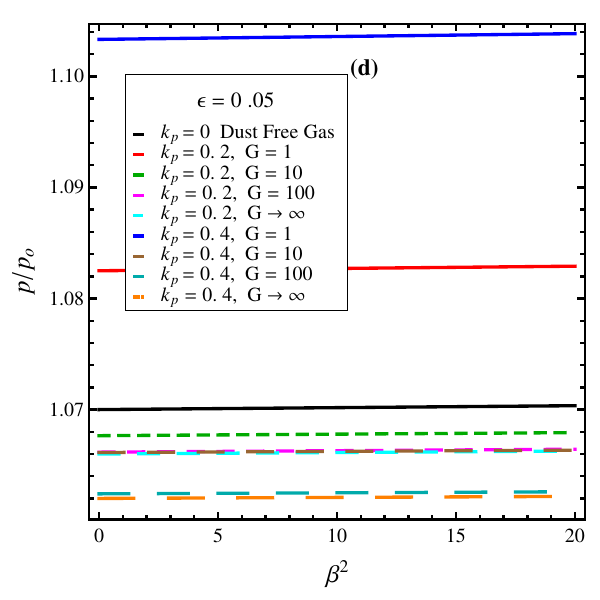}
     \end{minipage}
     \caption{Variations of (a) $U/a_{o}$ vs. $\xi$, (b) $p/p_{o}$ vs. $\xi$, (c) $U/a_{o}$ vs. $\beta^2$, (d) $p/p_{o}$ vs. $\beta^2$ in case of weak MHD shock waves in the presence of a strong magnetic field}\label{figure2a_2d}
\end{figure}
 
\subsubsection{Weak shock with weak magnetic field} 
The handy form of RH conditions for weak MHD shock waves in the presence of a weak magnetic field is given by equation (\ref{eq-3}). The shock velocity and the state variables are dependent on a parameter $\epsilon$, which is negligible in comparison with unity, the strength of magnetic field $\beta^2$, the mass concentration of solid particles $k_p$, the ratio of the density of solid particles to the initial density of gas $G$, the specific heat ratio of the solid particles $\beta_{sp}$, and the adiabatic index of gas $\gamma$. The numerical computations of the shock velocity and the pressure have been carried out taking the parameters as $0<\epsilon<0.1$, $0<\beta^2<1$, $k_p = 0, 0.2, 0.4$, $G = 1, 10, 100,\infty$, $\beta_{sp}=1$, and $\gamma =7/5$. The variations in the shock velocity and the pressure, respectively, with parameter $\epsilon$ and magnetic field $\beta^2$ for $\beta_{sp}=1$, $\gamma = 1.4$, and various values of $k_p$ and $G$ are shown in figure \ref{figure1a_1d}. It is important to note from figure \ref{figure1a_1d} (a)-(b) that the shock velocity and the pressure increase with the parameter $\epsilon$. The shock velocity increases with the parameter $k_p$ for the values of $G\leq 10$, whereas it decreases for $G=100$. An increase in the ratio of the density of solid particles to the initial density of gas $G$ leads to a decrease in the shock velocity. This behavior of the shock velocity, especially for the case of $k_p=0.4$ and $G=1$, differs greatly from the case of a dust-free (ideal) gas. The pressure with the parameter $k_p$ increases for $G = 1$; however, it decreases for the values of $G \geq 10$. An increase in the ratio of the density of solid particles to the initial density of gas $G$ leads to a decrease in the pressure with a constant value of $k_p$. This behavior of the pressure, especially for the case of $k_p=0.4$ and $G=1$, differs greatly from the case of a dust-free gas. It is also obvious from figure \ref{figure1a_1d} (c)-(d) that an increase in the strength of magnetic field $\beta^2$ leads to an increase in the shock velocity. However, the pressure is independent of the strength of the magnetic field.  

\subsubsection{Weak shock with strong magnetic field} 
The handy form of RH conditions for weak MHD shock waves in the presence of strong magnetic field is given by equation (\ref{eq-4}). The shock velocity and the state variables are dependent on a parameter $\epsilon$, the strength of the magnetic field $\beta^2$, the mass concentration of the solid particles $k_p$, the ratio of the density of solid particles to the initial density of gas $G$, the specific heat ratio of the solid particles $\beta_{sp}$, and the adiabatic index of gas $\gamma$. The numerical computations of the shock velocity and the pressure have been carried out taking the parameters as $0<\epsilon<0.1$, $0<\beta^2<20$, $k_p = 0, 0.2, 0.4$, $G = 1, 10, 100,\infty$, $\beta_{sp}=1$, and $\gamma =7/5$. The variations in the shock velocity and the pressure, respectively, with parameter $\epsilon$ and magnetic field $\beta^2$ for $\beta_{sp}=1$, $\gamma = 1.4$, and various values of $k_p$ and $G$ are shown in figure \ref{figure2a_2d}. It is obvious from figure \ref{figure2a_2d}(a)-(b) that the shock velocity and the pressure increase with the parameter $\epsilon$. The shock velocity with the parameter $k_p$  increases for the values of $G\leq 10$; however, it decreases for $G=100$. An increase in the ratio of the density of solid particles to the initial density of gas $G$ leads to a decrease in the shock velocity for a constant value of $k_p$. This behavior of the shock velocity, especially for the case of $k_p=0.4$ and $G=1$, differs greatly from the case of a dust-free gas. The pressure with the parameter $k_p$ increases for $G=1$; however, it decreases for the values of $G\geq 10$. An increase in the ratio of the density of solid particles to the initial density of gas $G$ leads to a decrease in the pressure for a constant value of $k_p$. This behavior of the pressure, especially for the case of $k_p=0.4$ and $G=1$, differs greatly from the case of a dust-free gas. Figure \ref{figure2a_2d}(c)-(d) shows that an increase in the strength of the magnetic field $\beta^2$ leads to an increase in the shock velocity. However, the pressure is independent of the strength of the magnetic field.   
\subsection{Strong MHD shock waves} 
Now we investigate the influence of the magnetic field on the shock velocity and the pressure across the strong MHD shock front in a two-phase gas-particle atmosphere under two conditions, viz., (i) when the magnetic field is weak and (ii) when the field is strong, respectively.

\begin{figure}   
   \begin{minipage}{0.4\textwidth}
     \centering
     \includegraphics[width=1.2\linewidth, trim=0 .1cm .1cm 0, clip]{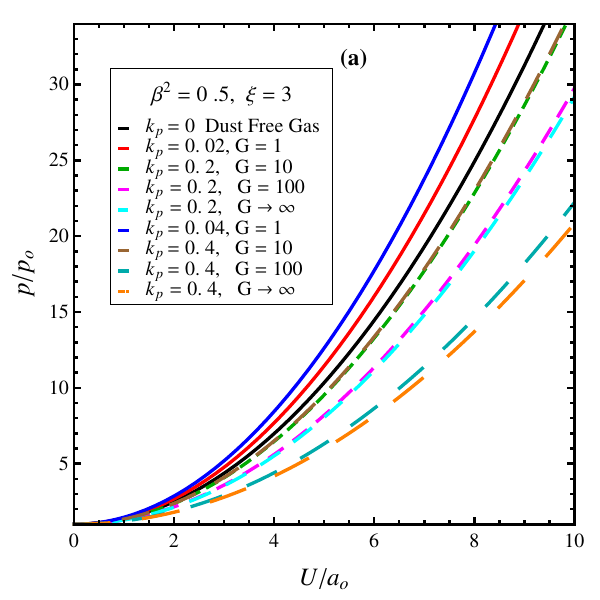}
       \end{minipage}\hfill
   \begin{minipage}{0.4\textwidth}
     \centering
     \includegraphics[width=1.2\linewidth, trim=0 .1cm .1cm 0, clip]{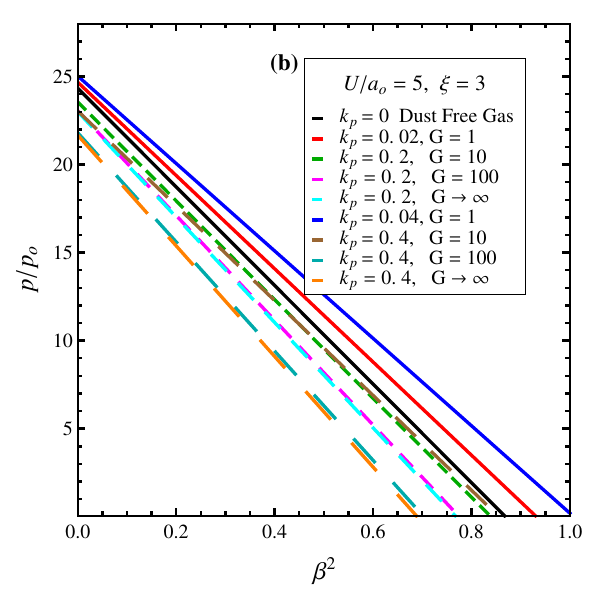}
     \end{minipage}
     \begin{minipage}{0.4\textwidth}
     \centering
     \includegraphics[width=1.2\linewidth, trim=0 .1cm .1cm 0, clip]{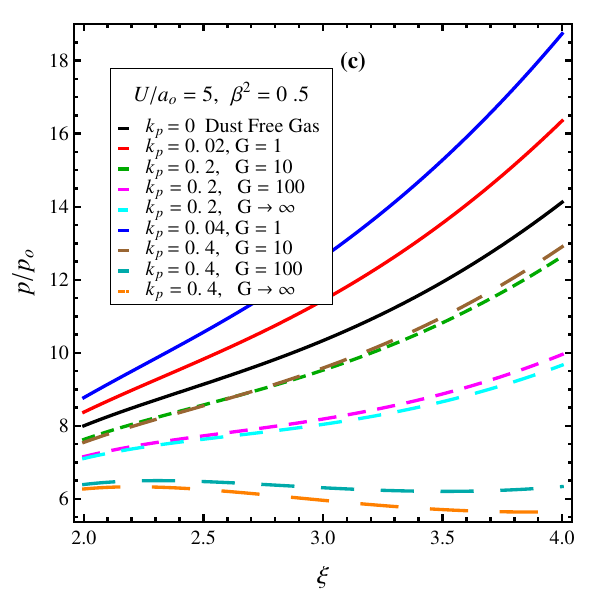}
       \end{minipage}\hfill
       \caption{Variations of (a) $p/p_{o}$ vs. $U/a_o$, (b) $p/p_{o}$ vs. $\beta^2$, (c) $p/p_{o}$ vs. $\xi$ in case of strong MHD shock waves in the presence of a weak magnetic field}\label{figure3a_3c}
\end{figure}
\begin{figure}   
   \begin{minipage}{0.4\textwidth}
     \centering
     \includegraphics[width=1.2\linewidth, trim=0 .1cm .1cm 0, clip]{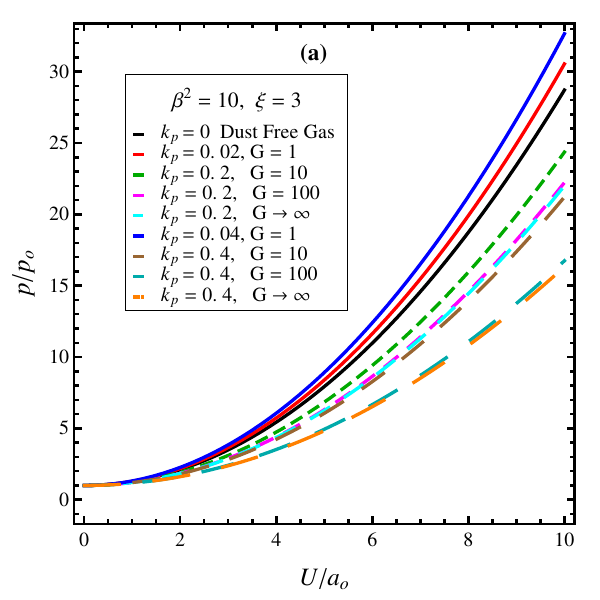}
       \end{minipage}\hfill
   \begin{minipage}{0.4\textwidth}
     \centering
     \includegraphics[width=1.2\linewidth, trim=0 .1cm .1cm 0, clip]{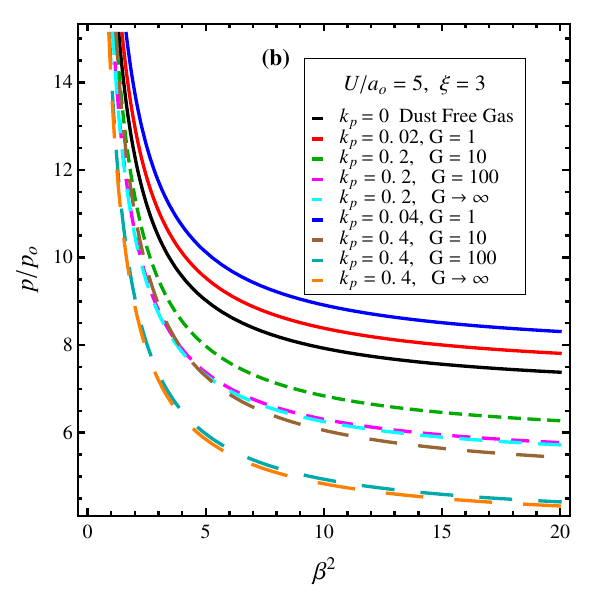}
     \end{minipage}
     \begin{minipage}{0.4\textwidth}
     \centering
     \includegraphics[width=1.2\linewidth, trim=0 .1cm .1cm 0, clip]{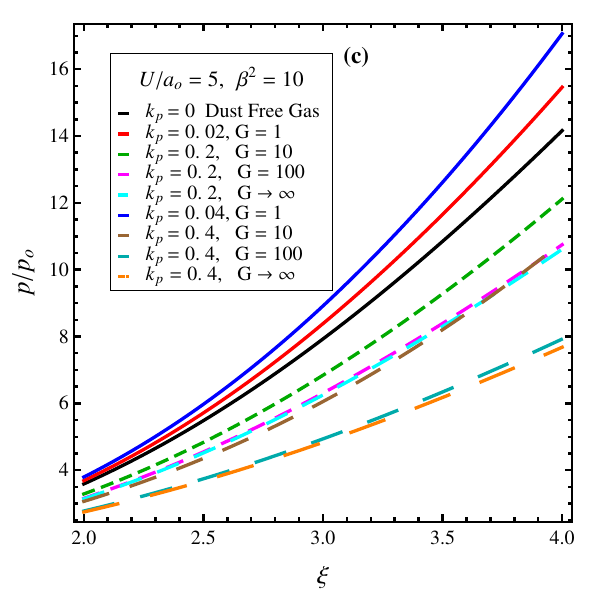}
       \end{minipage}\hfill
       \caption{Variations of (a) $p/p_{o}$ vs. $U/a_o$, (b) $p/p_{o}$ vs. $\beta^2$, (c) $p/p_{o}$ vs. $\xi$ in case of strong MHD shock waves in the presence of a strong magnetic field}\label{figure4a_4c}
\end{figure}

\subsubsection{Strong shock with weak magnetic field}
The handy form of RH conditions for strong MHD shock waves in the presence of a weak magnetic field is given by equation (\ref{eq-5}). The shock velocity and the state variables are dependent on a parameter $\xi$, which is known as the shock compression ratio, the strength of the magnetic field $\beta^2$, the mass concentration of the solid particles $k_p$, the ratio of the density of solid particles to the initial density of gas $G$, the specific heat ratio of solid particles $\beta_{sp}$, and the adiabatic index of gas $\gamma$. The numerical computations of the pressure have been carried out taking the parameters as $2<\xi<4$, $0<\beta^2<1$, $0<U/a_o<10$, $k_p =0, 0.02, 0.2, 0.4$, $G =1, 10, 100, \infty$, $\beta_{sp}=1$, and $\gamma =7/5$. The variations in the pressure, respectively, with the shock velocity $U/a_o$, the strength of the magnetic field $\beta^2$, and the shock compression ratio $\xi$ for $\beta_{sp}=1$, $\gamma = 1.4$, and various values of $k_p$ and $G$ are shown in figure \ref{figure3a_3c}. The pressure increases with the shock strength and the shock velocity; however, it decreases with the strength of the magnetic field. The pressure with the parameter $k_p$ increases for the value of $G=1$, whereas it decreases for $G\geq100$. An increase in the ratio of the density of solid particles to the initial density of gas $G$ leads to a decrease in the pressure for a constant value of $k_p$. This behavior of the pressure, especially for the case of $k_p=0.4$ and $G=1$, differs greatly from the case of a dust-free gas. It is notable that in the presence of a strong magnetic field, the trend of variation of the pressure across the strong MHD shock front in a dusty gas atmosphere is similar to that of across the strong MHD shock front in a dust-free gas.

\subsubsection{Strong shock with strong magnetic field}
The handy form of RH conditions for strong MHD shock waves in the presence of a strong magnetic field is  given by equation (\ref{eq-6}). The shock velocity and the state variables are dependent on a parameter $\xi$, the strength of the magnetic field $\beta^2$, the mass concentration of solid particles $k_p$, the ratio of the density of solid particles to the initial density of gas $G$, the specific heat ratio of solid particles $\beta_{sp}$, and the adiabatic index of gas $\gamma$. The numerical computations of the pressure have been carried out taking the parameters as $2<\xi<4$, $0<\beta^2<20$, $0<U/a_o<10$, $k_p =0, 0.02, 0.2, 0.4$, $G =1, 10, 100, \infty$, $\beta_{sp}=1$, and $\gamma =7/5$. The variations in the pressure, respectively, with the shock velocity $U/a_o$, the strength of magnetic field $\beta^2$, and the shock compression ratio $\xi$ for $\beta_{sp}=1$, $\gamma = 1.4$, and various values of $k_p$ and $G$ are shown in figure \ref{figure4a_4c}. The pressure increases with the shock velocity and the shock strength. It is obvious from figure \ref{figure4a_4c}(b) that the pressure first decreases rapidly and then becomes almost constant with the strength of the magnetic field $\beta^2$. The pressure with the parameter $k_p$ increases for the value of $G=1$, whereas it decreases for $G\geq 10$. An increase in the ratio of the density of solid particles to the initial density of gas $G$ leads to a decrease in the pressure for a constant value of $k_p$. This behavior of the pressure, especially for the case of $k_p=0.4$ and $G=1$, differs greatly from the case of a dust-free gas. It is notable that in the presence of a strong magnetic field, the trend of variation of the pressure across the strong MHD shock front in a dusty gas atmosphere is similar to that of across the strong MHD shock front in a dust-free gas.

\section{Conclusions}
\label{concl}
The present work has shown that the shock velocity and the pressure across the MHD shock front in a dusty gas atmosphere are mainly affected by the dust loading parameters and strength of the magnetic field. The velocity of weak shock increases with the strength of the magnetic field. However, the pressure across the weak shock front is independent of the magnetic field. The pressure across a strong MHD shock front increases with the shock velocity and the shock strength. However, it decreases with the strength of the magnetic field. The trends of variations of the shock velocity and the pressure across the MHD shock front in a two-phase gas-particle atmosphere are similar to those across the MHD shock front in a dust-free gas, in general.

\textbf{Acknowledgements} I acknowledge the support and encouragement of my family. 



\end{document}